\documentclass[aps,prl,amsmath,reprint,superscriptaddress,longbibliography]{revtex4-2} 
\usepackage{hyperref}
\usepackage{graphicx} 
\usepackage{units}
\usepackage{upgreek}
\usepackage{color}
\usepackage{amsmath,amssymb,amsfonts}
\begin{document}

%\title{Spontaneous folding transition in chiral active granular polymers}

\title{Spontaneous self-wrapping in chiral active polymers}

\author{Lorenzo Caprini}
\email{lorenzo.caprini@gssi.it} 
\affiliation{Institut f\"ur Theoretische Physik II: Weiche Materie, Heinrich-Heine-Universit\"at D\"usseldorf, D-40225 D\"usseldorf, Germany}
\affiliation{Physics Department, Sapienza University of Rome, Piazzale Aldo Moro, 5, 00185, Rome, Italy}

\author{Iman Abdoli}
\affiliation{Institut f\"ur Theoretische Physik II: Weiche Materie, Heinrich-Heine-Universit\"at D\"usseldorf, D-40225 D\"usseldorf, Germany}

\author{Umberto Marini Bettolo Marconi}
\affiliation{University of Camerino, via Madonna delle Carceri, 62032, Camerino, Italy}

\author{Hartmut L\"owen}
\affiliation{Institut f\"ur Theoretische Physik II: Weiche Materie, Heinrich-Heine-Universit\"at D\"usseldorf, D-40225 D\"usseldorf, Germany}

\date{\today}

\begin{abstract}
Biological organisms often have elongated, flexible structures with some degree of chirality in their bodies or movements. In nature, these organisms frequently  take advantage of self-encapsulation mechanisms that create folded configurations, changing their functionality, such as for defensive purposes. Here, we explore the role of chirality in polymeric structures composed of chiral active monomers exhibiting circular motion. Through a combination of experiments and numerical simulations, we demonstrate a spontaneous unfolding-folding transition uniquely induced by chirality, a phenomenon not observed in passive polymers. This transition is driven by a self-wrapping mechanism, resulting in dynamic polymer collapse even without attractive interactions. Our findings, based on chiral polymers made from chiral active granular particles, present new opportunities in robotic applications 
taking advantage of  the interplay between chirality and deformability.
\end{abstract}

\maketitle

%\section*{Introduction}

\noindent

\paragraph*{Introduction --} Chirality, the property of objects to be non-superimposable on their mirror images, is a fascinating and ubiquitous property in science, in particular in physics, chemistry, and biology.
From a fundamental perspective, chirality has been hypothesized and observed in particle physics: fermions and antifermions engaging in the charged weak interaction, are left-chiral and right-chiral, respectively: Because of their coupling, it has been postulated that the entire universe is left-handed chiral.
In chemistry, several molecules, such as the common sugar Glucose, are characterized by an intrinsic chirality~\cite{prelog1976chirality}. A similar scenario is encountered in biology, for instance in more complex, biological molecules, such as amino acids, and several cells~\cite{xu2007polarity}, where chirality arises from the self-organization of the actin cytoskeleton~\cite{tee2015cellular}.

In the realm of active matter systems~\cite{marchetti2013hydrodynamics, elgeti2015physics, bechinger2016active}, the interplay between particle motion and chirality has recently emerged as a captivating frontier, offering insights into an array of fascinating phenomena~\cite{liebchen2022chiral}. Chirality introduces an intriguing twist to the dynamics of active particles~\cite{lowen2016chirality}, leading to a unique class of emergent phenomena~\cite{zhang2020reconfigurable, siebers2023exploiting, liao2021emergent, ceron2023diverse, caprini2023chiral}, ranging from hyperuniform phases in dilute systems~\cite{lei2019nonequilibrium, huang2021circular, zhang2022hyperuniform} to microphase separation~\cite{liebchen2017collective} and self-reverting vorticity~\cite{caprini2024self} in denser systems.
The rotational symmetry breaking due to chirality induces the emergence of odd properties~\cite{fruchart2023odd}, such as odd elasticity~\cite{braverman2021topological, alexander2021layered, tan2022odd}, odd viscosity~\cite{banerjee2017odd, markovich2021odd, lou2022odd} and, recently, odd diffusivity~\cite{hargus2021odd}, which offer intriguing perspectives for the design of metamaterials~\cite{bertoldi2017flexible}.
Chirality has been experimentally observed in biological active matter systems consisting of elongated flexible bodies, such as cytoskeleton filaments~\cite{dunajova2023chiral} and protofilaments in bacteria cells~\cite{shi2020chiral}, while several bacteria themselves are characterized by a degree of chirality~\cite{woolley2003motility}.
Despite the huge theoretical work on active polymers~\cite{winkler2017active, anand2018structure, bianco2018globulelike, anand2020conformation, deblais2020phase, vliegenthart2020filamentous, kurzthaler2021geometric, fazelzadeh2023effects, faluweki2023active, janzen2024density,  fazelzadeh2023active, abbaspour2023effects}, i.e. filaments consisting of self-propelled particles, the role of chirality in these systems has been poorly explored.

\begin{figure*}
	\includegraphics[width=0.9\textwidth]{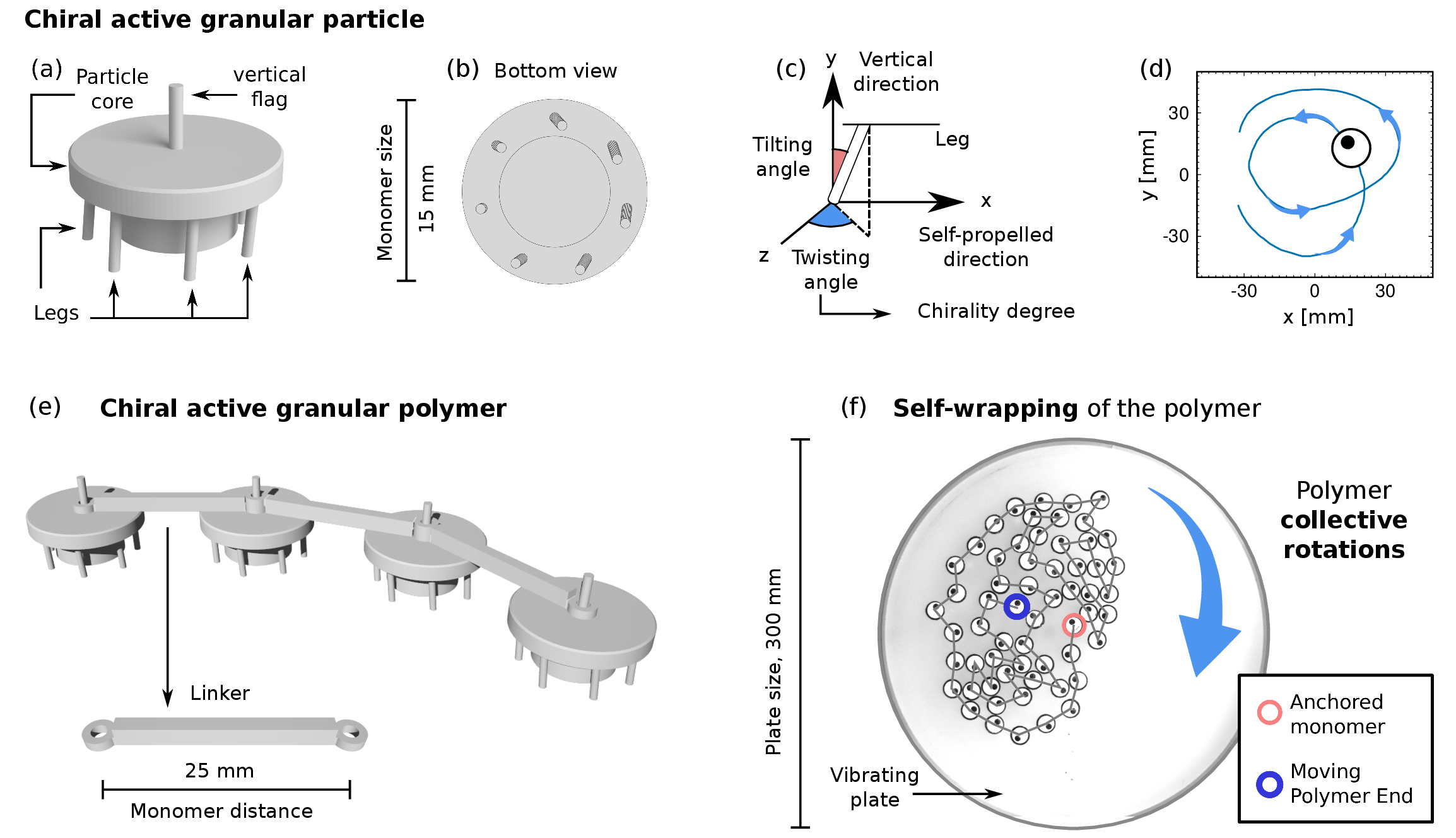}
	\caption{\label{fig:presentation}\textsf{\textbf{Chirality-induced self-wrapping and polymer folding.} 
	(a)-(b) 3D and bottom view of a chiral active granular particle.
(c) Illustration of the leg design showing both tilting and twisting angles.
(d) Time-trajectory of chiral active granular particles showing circular motion.
(e) Illustration of a chiral polymer consisting of chiral active granular particles kept together by a 3D-printed linker. 
(f) Folded configuration displayed by a chiral polymer anchored to the middle of the plate which is further characterized by a persistent global rotation, as suggested by the blue arrow). The anchored monomer is marked in red, while the free-to-move polymer end is marked in blue. 
	}}
\end{figure*}

Here, we discover a spontaneous folding transition in chiral active polymers, i.e.\ chain-like structures consisting of chiral active monomers. This transition is powered by a self-wrapping mechanism which is revealed by a finite average winding number. The self-wrapping process is responsible for the polymer collapse despite the absence of attractive interactions between the fundamental polymer units.
Our  results are obtained via active granular experiments~\cite{aranson2007swirling, kumar2014flocking, scholz2018rotating, koumakis2016mechanism, baconnier2022selective, lopez2022chirality, xu2024constrained, caprini2024emergent}. After designing and fabricating chiral active granular particles (Fig.~\ref{fig:presentation}~(a)-(c)), which exhibit circular trajectories (Fig.~\ref{fig:presentation}~(c)), we linked them together to create a chiral active granular polymer with a tunable structure. The experimental results presented here are supported by numerical simulations, which elucidate the fundamental role of chirality as the primary mechanism driving the spontaneous folding transition.

%Our findings highlight the crucial role of chirality in active polymer complex formation which induces topologically  locked-in  self-organization phenomena with nonequilibrium polymer scaling laws.
%Also various applications of the self-wrapping process are lying ahead: The topological-protected winding number can be used to store information in a fluctuating environment by active chirality. 
%The self-locking mechanism paves the way for the design of autonomous soft robots by taking advantage of the interplay between deformability and chirality.
Our findings highlight the crucial role of chirality in active polymer structure which induces self-organization phenomena with nonequilibrium polymer scaling laws. The self-wrapping mechanism paves the way for the design of autonomous soft, multicomponent robots with a degree of deformability. Rotations in the single components of a robot may be employed to enhance the cohesivity of the whole robotic structure.
These robotic agents may use self-wrapping in a biomimetic spirit exploiting the rolling-up strategy of pangolins and armadillos to protect themselves against predators.

%\newpage
%\section*{Results}

\paragraph*{Chiral active granular particles --} We design chiral active granular particles as 3D-printed plastic objects with broken translational and rotational symmetry.
These particles possess a cylindrical body (Fig.~\ref{fig:presentation}~(a)), and multiple legs that contact a vibrating baseplate.
When subjected to vertical vibrations generated by an electromagnetic shaker, these particles exhibit stochastic motion, the characteristics of which depend on the design of their legs.
Tilting all the legs in the same direction induces direct self-propelled motion at speed $v_0$ for a typical duration, defining the particle persistence time, $\tau$. Furthermore, twisting all the legs either clockwise or counterclockwise disrupts rotational symmetry (Fig.~\ref{fig:presentation}~(b)-(c)), resulting in a self-rotating motion with a typical angular velocity, $\omega$, often called chirality. The superposition of these effects enables the granular particles to execute circular trajectories (Fig.~\ref{fig:presentation}~(d)) with a typical radius $R \sim v_0/\omega$.
Therefore, our object behaves as  chiral active granular particles.
Further details on the particle chiral motion, such as oscillating mean-square displacement~\cite{van2008dynamics, reichhardt2019active} and long-time diffusion coefficient~\cite{van2008dynamics, caprini2019active}, are reported in the Supplemental Materials (SM).

\begin{figure}
	\includegraphics[width=\columnwidth]{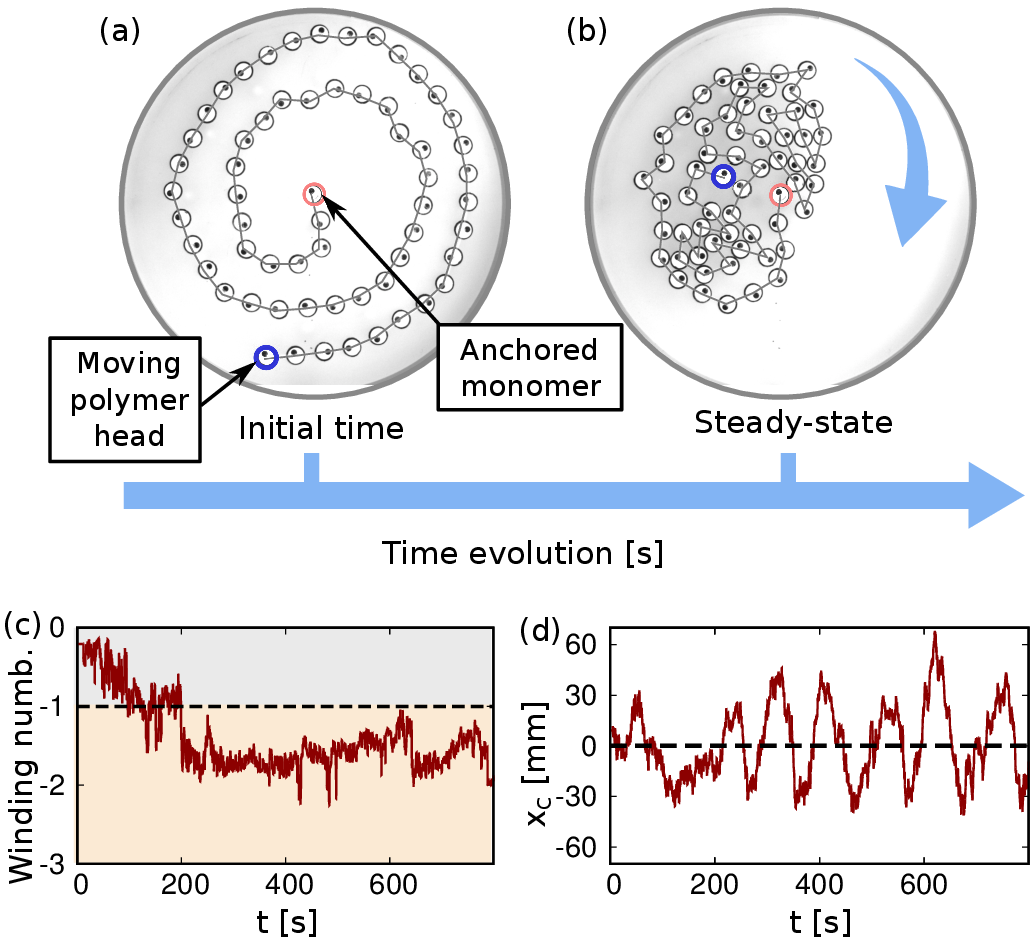}
	\caption{\label{fig:configuration}\textsf{\textbf{Self-trapping in spontaneous folded configurations.} 
	(a)-(b) Experimental snapshot configurations for different times until the steady state is reached. The polymer tail, placed in the middle of the plate, is anchored to the plate and is marked with a red circle, while the free-to-move polymer head is marked with a blue circle. The blue arrow suggests that the polymer performs collective rotations around the plate center. 
(c) Winding number relative to the free-to-move polymer tail as a function of time $t$, showing the polymer self-trapping. Here, a horizontal line marks the value $-1$ splitting configurations where the polymer tail is self-trapped (yellow region) and unrolled (grey region).
(d) center of mass time evolution of the polymer ($x$-coordiate) as a function of time, showing collective rotations of the polymer.
Measurements are obtained for a polymer with $N=60$ monomers.
	}}
\end{figure}

\begin{figure*}
	\includegraphics[width=\textwidth]{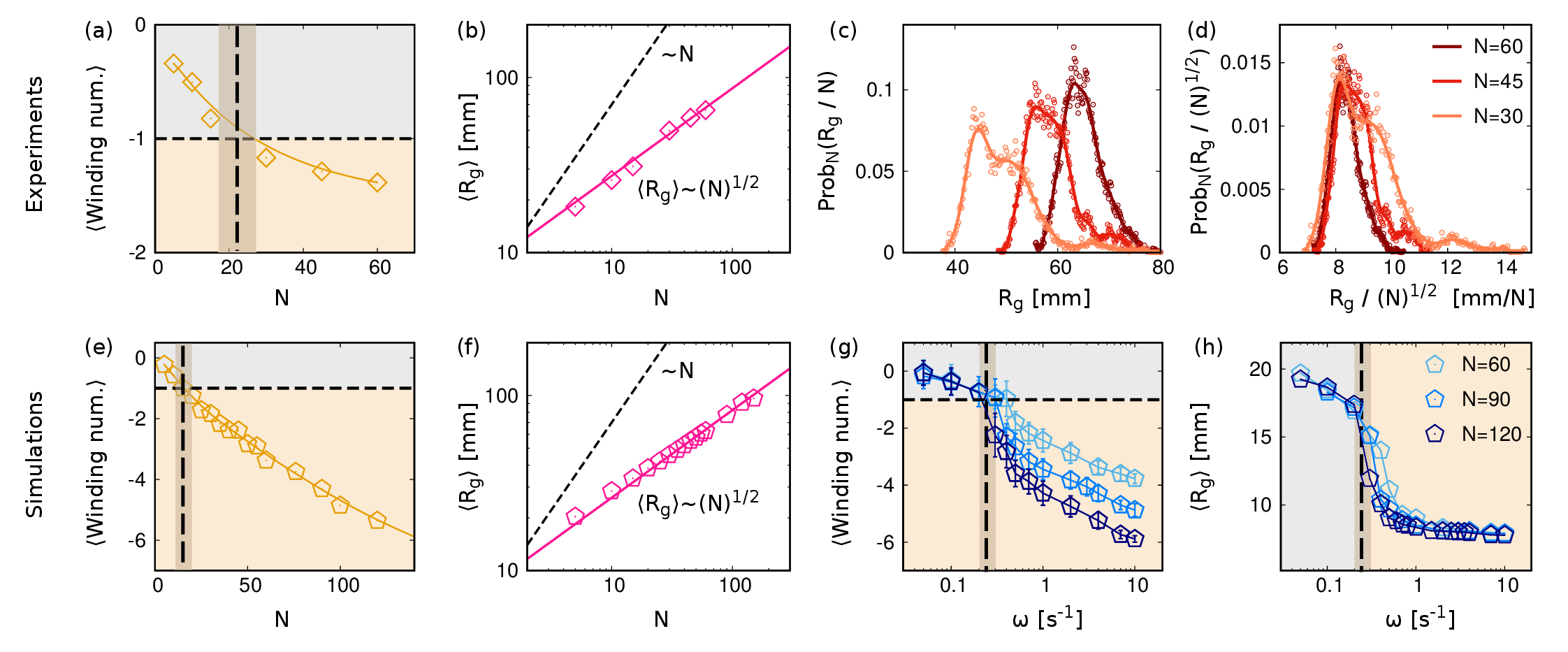}
	\caption{\label{fig:mean_observables}\textsf{\textbf{Chirality induced unfolding-folding transition.} 
	(a) and (e): Average winding number as a function of the polymer size, i.e.\ the number of monomers $N$, from experiments (a) and simulations (e). 
(b) and (f): Average gyration radius $\langle R_g \rangle$ as a function of $N$, from experiments (b) and simulations (f).
(c) Probability distribution, $\text{Prob}(R_g)$, of the polymer gyration radius $R_g$. (d) $\text{Prob}(R_g)$ for the rescaled gyration radius $R_g/N$. Both (c) and (d) are obtained for for $N=60, 45, 30$.
Here, solid pink lines fit the function $\sim N^{1/2}$, while the dashed black line provides the $\sim N$ scaling.
(g) Average winding number as a function of the chirality $\omega$. 
(h) $\langle R_g \rangle$ as a function of $\omega$. 
Both (g) and (h) are calculated from simulations.
Vertical dashed lines in (g) and (h) marked the critical value of $\omega$ where the folded-unfolded transition occurs.
%
%In all panels, solid lines are a guide for the eyes while points are obtain while the horizontal dashed line marks the value $-1$: 
The polymer head is self-trapped if the winding number is smaller than -1 (yellow region).
Errors in (a)-(b) and (f)-(h), calculated from the standard deviation, are smaller than the point size.
Simulations are run with the single monomer parameters $v_0= \unit[56]{mm\, s^{-1}}$ (self-propelled speed), $\gamma=\unit[27]{g\,s^{-1}}$ (translational friction coefficient), $D_t= \unit[3.75]{mm^2 s^{-1}}$ (translational diffusion coefficient), $D_r=\unit[0.72]{s^{-1}}$ (rotational diffusion coefficient), $\omega=\unit[3.99]{s^{-1}}$ (chirality), and $\gamma_r=\unit[244]{g\,mm^2 s^{-1}}$ (rotational friction coefficient). The parameter of the monomer-monomer interactions are $\sigma=\unit[15]{mm}$ (monomer diameter), $L=\unit[22]{mm}$ (length of the monomer link), $\epsilon=\unit[1]{g\,mm^2s^{-2}}$, and $k=\unit[10^3]{g\,mm^2s^{-2}}$.
	}}
\end{figure*}

%\vskip0.2cm
%\noindent
\paragraph*{Chiral active granular polymers --}
An experimental realization of a chiral active polymer is realized by linking the centers of chiral active granular particles with a rigid, plastic rod capable of free rotations (Fig.~\ref{fig:presentation}~(e)). This  configuration allows us to create a flexible chain of granular particles, each of which independently exhibits circular motion.

To avoid collisions with the system's boundary in experiments, we anchor the end of the polymer to the center of the plate %Every measurements and start with the polymer in an ordered, unrolled configuration with respect to the polymer head 
(Fig.~\ref{fig:configuration}~(a)).
At first, we observe that our polymer behaves overall as an elongated structure with an intrinsic degree of chirality.
This is evident by monitoring the time-trajectory of the whole polymer which persistently rotates around the anchored monomer: the center of mass position, specifically the $x$-component ($x_{cm}$), shows time oscillations as a function of time $t$ (Fig.~\ref{fig:configuration}~(d)). % and is confirmed for different polymer lengths (See SM), i.e.\ different number of monomers $N$.
This behavior can be explained by recognizing that the polymer's center of mass behaves as a chiral active particle, inherently characterized by a non-zero average angular velocity.

%\vskip0.2cm
%\noindent
\paragraph*{Spontaneous polymer folding --}
Even starting from unrolled initial configurations with respect to the free-to-move polymer head, the system spontaneously evolves toward a spontaneous folding state, where the polymer is permanently enrolled around its head (Fig.~\ref{fig:configuration}~(b)).
This steady-state configuration is induced by a self-wrapping mechanism, observed both experimentally and numerically:
In the folded configuration, the freely moving polymer head is completely surrounded by other monomers, which effectively induce a self-trapping mechanism.
These surrounding monomers hinder the motion of the polymer head, preventing it from easily escaping and unfolding the polymer. 
Consequently, once this configuration is reached, the polymer can be considered in a metastable state.
This behavior is confirmed for various polymer lengths (number of monomers 
$N$) and is independent of the initial configurations (see SM). This effect is entirely due to chirality: The circular trajectories do not easily allow the monomers to explore the space and find the correct path to unfold the polymer.

To quantify the polymer self-wrapping mechanism, we study the winding number, $\mathcal{W}$, calculated on the free-to-move polymer head (Fig.~\ref{fig:configuration}~(c)).
By describing the polymer as a continuous curve connecting the monomers, this observable counts the total number of loops performed counterclockwise around the polymer head (see Methods for the definition of $\mathcal{W}$).
At the initial time, when the polymer is unrolled with respect to the head, $\mathcal{W}(t) \approx 0$. After a transient period, $\mathcal{W}(t)$ reaches values smaller than $-1$, indicating that the polymer performs at least one loop around the head. 
In this case, self-wrapping is achieved and the polymer reaches a folded configuration.
This effect is not observed for small polymers, $N\leq15$ while it becomes stronger as the polymer size is increased until it approaches values close to $-2$ for $N=60$. 
This is shown by studying the average winding number $\langle\mathcal{W}\rangle$ as a function of the number of monomers $N$ (Fig.~\ref{fig:mean_observables}~(a)).

%\vskip0.2cm
%\noindent
\paragraph*{Chirality-induced unfolding-folding transition --}
 The mechanism that leads to folded configurations is purely based on chirality. This idea is confirmed through simulations using a minimal model, consisting of a chiral active polymer made up of $N$ monomers.
Each monomer behaves as a chiral active granular particle, exhibiting underdamped active dynamics in the presence of chirality (see Methods for details).
 Linked monomers interact via a strong harmonic potential with a finite resting length,  ensuring that the distance between them remains fixed. Additionally, excluded volume is modeled through a purely repulsive potential.
Simulations qualitatively agree with experiments, showing that the average winding number $\langle \mathcal{W}\rangle$ monotonically decreases with $N$ and self-wrapping is achieved for a polymer size $N\sim 20$ (Fig.~\ref{fig:mean_observables}~(e)).
The quantitative discrepancy between experiments and simulations (smaller values of $\langle \mathcal{W}\rangle$ for large $N$) can originate from dissipative effects during collisions or additional torques due to rotational friction taking place in experiments. However, simulations show that these interaction mechanisms are not crucial to observe self-wrapping.

In addition, numerical simulations are used to investigate the main mechanism behind our experimental findings. Specifically, we observe that an increase in chirality $\omega$ induces a folding transition
(Fig.~\ref{fig:mean_observables}~(g)) where we use the average winding number as an order parameter:
For chirality values below a critical threshold $\omega<\omega_c$, the steady-state average winding number $\langle \mathcal{W}\rangle$ is close to zero and the polymer remains unfolded. By contrast, for $\omega>\omega_c$, the average winding number is smaller than -1 and the polymer permanently reaches a folded configuration, as confirmed for different polymer lengths $N$.
For small chirality, the polymer can easily disrupt a folded configuration because the head persistently moves in the same direction, pushing other monomers away.
However, this disruption does not occur with large chirality, where the polymer head follows circular trajectories with a small radius, significantly reducing its effective persistence and thus its ability to push away other monomers.

%\vskip0.2cm
%\noindent
\paragraph*{Anomalous Flory exponent --}
To extract information about the polymer's internal structure, particularly its compactness, we measure the gyration radius, $R_g^2 = \sum_i |\mathbf{x}_i- \langle \mathbf{x}\rangle|^2$, where the average is taken over all monomers.
In polymer physics, it is known that the average gyration radius $\langle R_g \rangle$ scales with the polymer size (number of monomers $N$) following a power-law $R_g \sim N^\nu$ where $\nu$ is called the Flory exponent~\cite{de1979scaling}. This exponent primarily depends on the system's dimensions and, for passive polymers consisting of repulsive particles, it can be exactly calculated through the free energy minimization: In two dimensions, $\nu=3/4$.

The spontaneous folding observed experimentally results in a gyration radius governed by an anomalous Flory exponent,
 $\nu \approx 1/2$, despite the presence of excluded volume effects, i.e. pure repulsion between different monomers.
This information has been derived from calculating the steady-state average $\langle R_g\rangle$  (Fig.~\ref{fig:mean_observables}~(b)) and the distribution of the gyration radius, $\text{Prob}(R_g)$ (Fig.~\ref{fig:mean_observables}~(c)). In particular, the distribution $\text{Prob}(R_g)$ obtained for different $N$ collapses by rescaling $R_g \to R_g/N^{1/2}$ (Fig.~\ref{fig:mean_observables}~(d)). 
This result is confirmed by numerical simulations where the exponent $\nu=1/2$ remains valid for larger polymer sizes $N$ (Fig.~\ref{fig:mean_observables}~(f)).
Interestingly, we note that $\nu=1/2$ in two dimensions is consistent with the collapse exponent, confirming the spontaneous folding observed in both experiments and simulations.
As a result, the average gyration radius can be used as an alternative order parameter to quantify the chirality-induced folding transition.
Indeed, $\langle R_g \rangle$ as a function of $\omega$ displays a sudden jump from large values where the polymer is unfolded to a small plateau value that is close to the minimal size compatible with a folded configuration (Fig.~\ref{fig:mean_observables}~(e)).

\paragraph*{Discussion --}
Here, the interplay between chirality and body flexibility is investigated by experimentally studying a granular polymer consisting of chiral active particles.
We discover that chirality significantly alters the polymer's conformational properties, leading to a spontaneous folding that has no equilibrium counterpart, as supported by numerical simulations.
The intrinsic chirality often found in natural polymer units can explain the folding of proteins or more complex biological filaments, as well as self-incapsulation mechanisms for defense purposes.

Our results challenge the conventional polymer theories applicable in equilibrium conditions, by showing that anomalous values of the Flory exponent can spontaneously arise purely as a non-equilibrium dynamical effect.
 The presence of chiral activity drives the polymer far from equilibrium, rendering its dynamics indescribable by free energy considerations.
Even with purely repulsive interactions, it is possible to observe small Flory exponents that are indicative of a folded polymer configuration.
Moreover, the spontaneous emergence of collective rotations opens new possibilities for designing efficient particle-based micro-motors. Indeed, previous studies have shown that bacteria and colloids can power asymmetric gears~\cite{hiratsuka2006microrotary, di2010bacterial, vizsnyiczai2017light}. The rotations induced by chirality may enhance motor efficiency, inspiring a future generation of micromotors.

%Unlike the asymmetric gears powered by active particles, such as bacteria or colloids, which have been studied previously~\cite{hiratsuka2006microrotary, di2010bacterial, vizsnyiczai2017light}, chirality may enhance motor efficiency.  This concept could inspire future experiments where chirality can be twisted.

%\section*{Acknowledgments} 
\paragraph*{Acknowledgments --} 
LC and HL thanks Olivier Dauchot for useful discussion.
LC acknowledges the European Union MSCA-IF fellowship for funding the project CHIAGRAM. 
IA and HL acknowledge support by the Deutsche Forschungsgemeinschaft (DFG) through the SPP 2265 under the grant number LO 418/25.

\bibliographystyle{naturemag}
\bibliography{bib}

%\section*{Methods}
\appendix

\section*{Appendices}

\vskip0.2cm
\noindent
\paragraph*{Experimental setup --}
To conduct our experiment, we employed a standard vibrating table, a well-established tool for inducing vibrational excitations in granular materials. 
Our setup comprises an acrylic baseplate with a diameter of $\unit[300]{mm}$ and a height of $\unit[15]{mm}$, complemented by an outer plastic ring that confines the particles within.
Vertical vibrations are generated by an electromagnetic shaker affixed to the acrylic baseplate. Horizontal alignment was meticulously adjusted to minimize the influence of gravitational drift. Additionally, we placed the setup on a substantial concrete block to counteract resonance effects with the environment.
Our choice of shaker frequency $f$ and amplitude $A$ was critical to achieving stable excitation without particle spillage. We operated at $f=\unit[120]{Hz}$ with an amplitude of $A=\unit[24(1)]{\upmu m}$, ensuring quasi-two-dimensional particle motion on the baseplate.

\vskip0.2cm
\noindent
\paragraph*{Tracking --}
We employed a high-speed camera, recording the system at a rate of $50$ frames per second with a resolution of $1024\times1024$ pixels. To determine the positions, $\mathbf{x}$, and orientations, $\theta(t)$, of the vibrobots, we utilized standard feature recognition methods, specifically the circle Hough transform, complemented by a custom classical algorithm designed for precise sub-pixel localization.
The process of reconstructing  particle trajectories involved tracking nearest neighbors between a pseudo-frame, extrapolated from the known positions and velocities of the previous frame, and the current frame. Translational and angular velocities $\mathbf{v}(t)$ and $\omega(t)$ were subsequently calculated from the displacements using the centered finite difference formulas: $\mathbf{v}=(\mathbf{x}(t+\Delta t)-\mathbf{x}(t-\Delta t))/(2\Delta t)$ and $\mathbf{\omega}=(\mathbf{\theta}(t+\Delta t)-\mathbf{\theta}(t-\Delta t))/(2\Delta t)$, where the time interval  $\Delta t=\unit[0.0066]{s}$. It is important to note that although these velocities correspond to finite displacements, they provide a sufficiently accurate approximation of the particles' instantaneous velocities, given the relatively large inertial relaxation time compared to the data acquisition time resolution.

\vskip0.2cm
\noindent
\paragraph*{Active chiral particle design --}
The particles used in our study are three-dimensional plastic objects created through 3D printing~\cite{scholz2018inertial}. Each particle possesses cylindrical symmetry and consists of a main body with multiple laterally attached legs.
The particle body  is composed of two primary components:
i) A cylindrical core with a diameter of $\unit[9]{mm}$ and a height of $\unit[4]{mm}$, designed to stabilize the particle by lowering its center of mass vertically.
ii) A larger cylindrical cap, with a diameter of $\unit[15]{mm}$ and a height of $\unit[2]{mm}$, which defines the particle's circular horizontal cross-section. This cap serves as the attachment point for seven cylindrical legs, each with a diameter of $\unit[0.8]{mm}$. These legs are regularly distributed around the core, forming a heptagonal configuration. The legs make contact with the ground, resulting in a total particle height of $\unit[7]{mm}$, with a vertical leg height of $\unit[5]{mm}$, and the core suspended $\unit[4]{mm}$ above the ground.
On average, each particle has a mass of $m=\unit[0.83]{g}$, and its moment of inertia, determined by the particle's shape, is $J=\unit[17.9]{g,mm^2}$.

To induce self-propelled, directed motion, we introduce a disruption in the particle's translational symmetry by tilting all the legs at a constant angle, $\alpha$, in the same direction. This angle governs the particle's polarization and is indicated by a white spot attached to the particle's cap. A larger $\alpha$ corresponds to a higher typical particle velocity. We set $\alpha$ to $\unit[4]{degrees}$ relative to the surface normal. Self-rotational motion is achieved by perturbing the cylindrical symmetry of the particle's legs. In this case, all legs are additionally twisted at the same angle relative to the vertical, counterclockwise direction (so that the particle rotates counterclockwise). A larger twisting angle results in a higher particle angular velocity and thus circular trajectories with a smaller radius. Our particle is characterized by a twisting angle of $\unit[3.4]{degrees}$.
Circular motion is attained  by simultaneously tilting and twisting the legs, combining translational and rotational symmetry-breaking mechanisms.

\vskip0.2cm
\noindent
\paragraph*{Dynamics of a chiral active granular particle --}
We characterize the dynamics of particles through a parametric approach based on modeling their motion using stochastic models for the particle dynamics. 
Specifically, the particle dynamics are described by the inertial active Brownian particle model, which encompasses underdamped dynamics for translational and rotational motion, which are coupled by the particle activity. The translational motion of an active vibrobot, with mass $m$, position $\mathbf x$, and velocity $\mathbf v=\dot{\mathbf x}$, is well-represented by a two-dimensional stochastic differential equation that balances inertial, dissipative, and active forces:
\begin{equation}\label{eq:dynamics_translational}
	m\dot{\mathbf v}+\gamma\mathbf v=\gamma\sqrt{2D_\text{t}}\boldsymbol\xi+\gamma v_0\mathbf n\,.
\end{equation}
Here, $\boldsymbol\xi$ is Gaussian white noise with zero average and unit variance, while $\gamma$ and $D_\text{t}$ describe the friction and the effective translational diffusion coefficients, respectively. The active driving force is defined by $\gamma v_0\mathbf n$, where $v_0$ represents the velocity mode, and $\mathbf n=(\cos\theta,\sin\theta)$ is the orientation vector based on the orientational angle $\theta$.

The rotational dynamics of a particle, characterized by its moment of inertia $J$, is governed by an additional equation of motion for the angular velocity $\Omega=\dot\theta$
\begin{equation}\label{eq:dynamics_rotational}
	J\dot\Omega=-\gamma_\text{r}\Omega+\gamma_\text{r}\sqrt{2D_\text{r}}\eta + \gamma_r\omega\,.
\end{equation}
Here, $\eta$ represents Gaussian white noise with zero average and unit variance, and $\gamma_\text{r}$ and $D_\text{r}$ are the rotational friction and rotational diffusion coefficients, respectively. Finally, the angular drift $\omega$ determines the particle chirality and represents the typical angular velocity of the chiral active particle.

\vskip0.2cm
\noindent
\paragraph*{Chiral active polymer design --}
Our chiral active particles feature a vertical central bar with a diameter of $0.1$ cm, which resembles a slender rod. These bars 
 are designed to facilitate the connection of two or more chiral active granular particles. This connection is achieved using 
 3D-printed rods, approximately $2.2$ cm long, with a thickness and height of $0.3$ cm. These plastic linkers are perforated at their ends, allowing them to be inserted into the vertical bars; the internal hole of these linkers has a radius of 0.2 cm.
By connecting two chiral active granular particles, we can form a dimer. Connecting two or more dimers together enables the creation of a polymer consisting of $N$ monomers.

\vskip0.2cm
\noindent
\paragraph*{Dynamics of a chiral active polymer --}
With this setup, the friction between slender rods and particles is negligible,
allowing the link we established to be effectively approximated solely as an interparticle force.
This implies that the orientation of each monomer can be described Eq.\eqref{eq:dynamics_rotational} without any torque interactions.
By contrast, the dynamics of the center of mass of the monomers are governed by Eq.\eqref{eq:dynamics_translational} with an additional interaction force $\mathbf{F}_i$, so that
\begin{equation}\label{eq:dynamics_translational}
m\dot{\mathbf v}_i+\gamma\mathbf v_i=\gamma\sqrt{2D_\text{t}}\boldsymbol\xi_i+\gamma v_0\mathbf n_i+\mathbf{F}_i\,,
\end{equation}
with $i=1, ..., N$.
The interparticle force is obtained by an interaction potential $\mathbf{F}_i=-\nabla_i U_{tot}$, which models the volume exclusion effect between different particles and the fixed distance between the particle centers
\begin{equation}
U_{tot}=  U_{ex} + U_{link} \,.
\end{equation}
Here, the exclusion is modeled by the potential
\begin{equation}
U_{ex} = \sum_{i<j}^N U_{WCA}
\end{equation}
where $U_{WCA}$ is the WCA potential given by
\begin{equation}
U_{WCA} = 4\epsilon \left( \left(\frac{\sigma}{r}\right)^{12} - \left(\frac{\sigma}{r}\right)^{6} \right)\,.
\end{equation}
The parameter $\sigma=15$ mm corresponds with the particle diameter while $\epsilon=\unit[1]{g \,mm^2s^{-2}}$ determines an energy scale and is large in numerical simulations. 
The inter-monomer links are modeled by  the following potential
\begin{equation}
U_{link} = \frac{k}{2\sigma^2} \sum^{N-1}_{i=1} \left(|\mathbf{x}_{i+1}-\mathbf{x}_i | - L \right)^2 \,,
\end{equation}
where $L=\unit[22]{mm}$ is the distance between neighboring monomers and $k$ the constant of this potential.
This parameter is set large enough so that the distance between neighboring monomers has negligible fluctuations, such that $k=\unit[10^3]{g \,mm^2s^{-2}}$.

\vskip0.2cm
\noindent
\paragraph*{Estimate of the particle parameters --}
The interaction potentials' parameters are selected to ensure that $\sigma$ 
represents the nominal particle diameter and to prevent large distance fluctuations between connected monomers.
In contrast, the intrinsic parameters of the individual chiral active particle are determined using a fitting algorithm that assesses translational and angular velocity distributions, mean squared displacement, and mean angular displacement.
Experimental outcomes are compared with simulation results obtained from
 Eqs. \eqref{eq:dynamics_translational} and \eqref{eq:dynamics_rotational}, iteratively adjusting parameters such as $v_0$, $\gamma$, $\gamma_r$, $D_r$, $D_t$, and $\omega$. 
Initially, these observables are evaluated with a predefined parameter set.
Subsequently, a Nelder-Mead optimization scheme is employed to minimize the overall deviation from the experimental data, iteratively searching for the optimal parameter set.
Our estimate of the model parameters is reported in the Table below:
\begin{center}
\label{tab1} 
\begin{tabular}{ c | c }
$v_0$ & 56.2231540 (4) mm s$^{-1}$  \\  
\hline
 $\gamma$ & 27.4532786 (2) g s$^{-1}$\\
\hline
$D_t$ & 3.75165050 (0.2) mm$^2$ s$^{-1}$ \\
\hline
 $D_r$ & 0.718160256 (0.01) s$^{-1}$ \\
\hline
$\omega$ & 3.98609903 (0.05) s$^{-1}$ \\ 
\hline
$\gamma_r$ & 244.332844 (30) g mm$^2$ s$^{-1}$ \,.
\end{tabular}
\end{center}
We observe that at the shaker conditions considered, the persistence time $1/D_r= \unit[1.4]{s}$ is larger than the other typical times governing the dynamics i.e. the chiral time $1/\omega = \unit[0.25]{s}$, the inertial time $m/\gamma=\unit[0.030]{s}$ and the rotational inertial time $J/\gamma_r=\unit[0.73]{s}$. Therefore, inertia does not play a fundamental role for the experimental setup of this paper.

\vskip0.2cm
\noindent
\paragraph*{Winding number calculation --}
We compute the winding number for each frame, which indicates the number of complete rotations particles make around the last particle, treated as the reference or source point. The angle of each particle relative to this last particle is calculated and then unwrapped to ensure continuity, avoiding sudden changes caused by the transition from
$-\pi$ to $\pi$. Then we compute the differences between consecutive angles and ensure that these differences do not exceed a predefined threshold. If they do, they are capped to prevent extreme jumps, which helps stabilize the calculation. The sum of these adjusted angle differences, divided by $2\pi$, gives the winding number for each frame, indicating the total number of rotations made by the chain around the last particle, normalized to the number of complete rotations.

%\section*{Author contributions}
%LC, UMBM, and HL conceived the project.
%LC designed and performed the experiment.
%IA and LC performed the experimental analysis and perform numerical simulations.
%LC wrote the first draft of the paper but all the authors equally contribute to the manuscript writing. 

%\section*{Competing financial interests}
%The authors declare no competing financial interests.

%\section*{Data availability}
%The data that support the plots within this paper and other findings of this study are available from the corresponding author upon request.

%\section*{Code availability}
%An STL file is included as Supplementary Information to reproduce the design of chiral active granular particles forming the chiral polymer.
%OpenSCAD file of the rotor design is included as Supplementary Information.

\end{document}